# Light bullets by synthetic diffraction-dispersion matching

Valery E. Lobanov,* Yaroslav V. Kartashov, and Lluis Torner

*ICFO-Institut de Ciencies Fotoniques, and Universitat Politecnica de Catalunya, Mediterranean Technology Park, 08860 Castelldefels (Barcelona), Spain*

We put forward a new approach to generate stable, fully three-dimensional light bullets, which is based on the matching of the intrinsic material dispersion with a suitable effective diffraction. The matching is achieved in adequate waveguide arrays whose refractive index is periodically modulated along the direction of light propagation. We show that by using non-conventional, out-of-phase longitudinal modulation of the refractive index of neighboring channels, it is possible to tune the effective diffraction to match the intrinsic material group velocity dispersion. Three-dimensional light bullets are shown to form at reduced energy levels, in settings where the dispersion would be far too weak to generate bullets in the absence of array.

*PACS numbers: 42.65.Tg, 42.65.Jx, 42.65.Wi.*

The concept of self-trapping of light in nonlinear media was introduced almost five decades ago [1]. Shortly after, Zabusky and Kruskal coined the term "soliton" for self-sustained excitations that propagate undistorted in nonlinear media and that experience elastic collisions [2]. Since then, a phenomenal progress has been achieved in the generation of solitons in different geometries and in materials with various nonlinearity mechanisms. Solitons may form in both one- and multidimensional settings. The term light bullet, i.e., a fully three-dimensional optical soliton localized in space and time was introduced almost three decades ago [3]. Such states form when nonlinearity arrests both spatial diffraction and temporal group velocity dispersion and they may find applications in future ultrafast optical processing devices with terahertz switching rates where each bullet represents a bit of information [4,5]. In principle, light bullets may be supported by a variety of nonlinear materials [6,7]. A different type of light bullets, in the form of Airy-Bessel pulses-beams propagating in linear media, has been recently realized in a landmark advance [8]. However, formation of stable fully three-dimensional light bullet in nonlinear media remains one of he most challenging and fascinating open problems in nonlinear optics.

Among the main obstacles preventing the experimental formation of light bullets is their instability in several nonlinear materials (including uniform media with cubic nonlinearity) and, most importantly, the lack of natural materials where nonlinearity, diffraction, and dispersion act on similar footing. To date only two-dimensional bullet formation was achieved in quadratic nonlinear media by generating dispersion via achromatic-phase-matching at the expense of one spatial dimension [9,10]. Thus, elucidation of new strategies that overcome the intrinsic material limitations and that enable to match the effective diffraction and dispersion strengths are of paramount importance. Several approaches were suggested, including utilization of so-called tandem structures where nonlinearity and dispersion are contributed by different materials arranged into structures that may be periodic in longitudinal [11] or transverse [12] directions and where light propagation depends on averaged characteristics of such medium.

In this Letter we introduce a novel approach that relies on engineering the strength of diffraction in suitable longitudinally modulated two-dimensional waveguide arrays [13-15]. One of the most remarkable features of longitudinally modulated waveguide arrays is the possibility of discrete diffraction suppression (or light tunneling inhibition), even in the linear regime. Such suppression is a resonant effect, occurring only for a specific set of amplitudes and frequencies of longitudinal refractive index modulation, as it was demonstrated in periodically curved arrays [16-20], in arrays with oscillating widths of channels [21] and oscillating refractive index [22-25]. Remarkably the effective diffraction strength in such structures is determined by the detuning of the modulation frequency from its resonant value; hence by changing the detuning one can reduce diffraction strength dramatically and even make it comparable with the strength of the intrinsic GVD. Thus, note that the concept is based on a fundamental change of approach. Namely, a material limitation (i.e., finding a natural material with intrinsic suitable local properties) is partially transformed into a design challenge (i.e., finding a synthetic structure with suitable overall properties). In contrast to the former, the latter is externally controllable.

We illustrate the power of the concept by analyzing light pulse propagation in two-dimensional honeycomb waveguide arrays with out-of-phase periodic longitudinal refractive index modulation in neighboring channels. We show that by varying the modulation frequency detuning one can engineer the effective diffraction in such a way that for a fixed pulse duration it is possible to select a detuning and an input beam amplitude in order to achieve the efficient excitation of stable three-dimensional light bullets, even when the material intrinsic group velocity dispersion (GVD) is much smaller than diffraction in the ab-sence of the array. As an aside consequence, such bullets form at reduced input light energies. Discrete mobile bullets in curved waveguide arrays are shown to exist in related geometries indeed [26]. However, note that such bullets extend over multiple channels in sharp contrast to the strongly localized states that we address here.

Our model is based on the nonlinear Schrödinger equation for the dimensionless field amplitude $q$, governing the propagation of spatio-temporal wave packet along the $\xi$ axis of the waveguide array with a longitudinally modulated refractive index:

$$i\frac{\partial q}{\partial \xi} = -\frac{1}{2}\left(\frac{\partial^2 q}{\partial \eta^2} + \frac{\partial^2 q}{\partial \zeta^2}\right) - \frac{1}{2}\beta\frac{\partial^2 q}{\partial \tau^2} - pR(\eta,\zeta,\xi)q - |q|^2 q. \tag{1}$$

Here $q = (\pi L_{\text{diff}} n_2 / \lambda)^{1/2} E$, $E$ is the field amplitude, $n_2$ is the nonlinearity coefficient, $L_{\text{diff}} = k_0 r_0^2$ is the diffraction length in a homogeneous medium, $k_0 = 2\pi n_0 / \lambda$ is the wavenumber, $n_0$ is the unperturbed refractive index at the wavelength $\lambda$, $\eta = x / r_0$, $\zeta = y / r_0$ and $\xi = z / L_{\text{diff}}$ are the transverse and longitudinal coordinates normalized to the characteristic transverse scale $r_0$ and diffraction length $L_{\text{diff}}$, respectively, $\tau = t / t_0$ is the normalized (retarded) time, $\beta = L_{\text{diff}} / L_{\text{disp}}$, $L_{\text{disp}} = |\partial^2 k / \partial \omega^2|^{-1} t_0^2$ is the dispersion length, and $p = k_0^2 r_0^2 \delta n / n_0$ describes the refractive index contrast in the individual waveguides. The refractive index distribution is represented by the function $R(\eta,\zeta,\xi)$ describing the array of Gaussian waveguides $\exp[-(\eta^2 + \zeta^2)/d^2]$ with centers $(\eta_k, \zeta_k)$ placed in the nodes of a honeycomb grid. We set $d = 0.5$ and $s = 2$, respectively. The refractive index in neighboring waveguides is modulated out-of-phase along $\xi$ axis, i.e., if in central waveguide the refractive index oscillates as $1 + \mu \sin(\Omega\xi)$, where $\mu \ll 1$, in all adjacent waveguides it changes as $1 - \mu \sin(\Omega\xi)$ [see Fig. 1(a) showing refractive index distribution in such an array at the distance $\xi = \pi / 2\Omega$]. We further set $\mu = 0.15$ and $p = 12$ and concentrate on the impact of the modulation frequency $\Omega$ on wave packet dynamics. We also suppose that the strength of the intrinsic GVD of the material is small compared to diffraction strength and set $\beta = 0.1$.

The out-of-phase modulation of the refractive index in neighboring guides allows engineering of diffraction in such array. In the absence of dispersion ($\beta = 0$) one can find that diffraction is almost completely inhibited even for linear single-channel excitations for certain resonant modulation frequencies [15]. We are interested only in primary highest-frequency resonance and denote corresponding frequency as $\Omega_{\text{r}}$. Figures 1(b) and 1(c) show linear output patterns at $\xi = T_{\text{b}} / 2$ (here $T_{\text{b}} = 67.17$ is the period of the energy beating between two unmodulated guides) in unmodulated array and modulated array with $\Omega = \Omega_{\text{r}}$ [in our case $\Omega_{\rm r} \approx 9.61(2\pi/T_{\rm b})$], respectively. One can see that while in the unmodulated array the diffraction is considerable, in the modulated array the light remains in the central waveguide. The "effective strength of diffraction" in this system is characterized by the detuning of modulation frequency from the resonant one. This is apparent from the dependence of the ratio of output $R_{\rm out}$ and input $R_{\rm in}$ integral beam radii on frequency detuning $(\Omega-\Omega_{\rm r})/\Omega_{\rm r}$ at the distance $\xi=2T_{\rm b}$ depicted in Fig. 2(a).

Engineering of diffraction by the frequency detuning drastically affects the conditions for the formation of light bullets when $\beta \neq 0$. In order to investigate full spatiotemporal dynamics we solved Eq. (1) with the input conditions in the form $q = Aw(\eta,\zeta)\,{\rm sech}(\nu\tau)$, where $w(\eta,\zeta)$ represents the profile of the linear mode of an isolated waveguide, $A$ is the input amplitude, while parameter $\nu$ characterizes input pulse duration. Further we set $\nu = 0.8$. In all cases considered the propagation distance was $L = 4T_{\rm b}$. To characterize the efficiency of the spatiotemporal localization we introduce a distance-averaged energy fraction trapped in central excited channel and within time interval $(-\tau_0,+\tau_0)$, where $\tau_0 = 1.657\nu^{-1}$ (this corresponds to time where amplitude in the input wave packet decreases $e$ times):

$$U_{\rm m} = L^{-1}\int_0^L d\xi \int_{-\tau_0}^{\tau_0} d\tau \int\int_{-s/2}^{s/2} |q(\eta,\zeta,\tau,\xi)|^2 \, d\eta d\zeta \bigg/ \int_{-\tau_0}^{\tau_0} d\tau \int\int_{-s/2}^{s/2} |q(\eta,\zeta,\tau,0)|^2 \, d\eta d\zeta. \quad (2)$$

In order to find the optimal amplitude for the excitation of three-dimensional light bullets we calculated the dependencies of the distance-averaged energy fraction in central channel on the amplitude of the input beam for different negative values of frequency detuning [Fig. 2(b)]. One of the central results of this Letter is that for each frequency detuning (effective diffraction strength) for the selected input pulse duration there exists an optimal input amplitude $A_{\rm opt}$ corresponding to the maximum in the dependence $U_{\rm m}(A)$ and most efficient excitation of light bullet (i.e., situation when radiative losses are minimal). Moreover, the best spatiotemporal localization (i.e. highest value of distance-averaged energy) is achieved when modulation frequency corresponds to the resonant one as it is seen from comparison of curves 1 and 2 in Fig. 2(b). For each frequency detuning the efficient spatiotemporal localization, or excitation of a light bullet, is possible when the input amplitude $A$ is sufficiently close to the optimal one. The distance-averaged energy decreases rapidly with decrease of input amplitude, but what is surprising is that it also decreases for sufficiently high $A$ values (the latter may be a consequence of the radiative losses and the nonlinearity-induced delocalization that could take place in longitudinally modulated systems). Importantly, for $\Omega = \Omega_r$ and for moderate detunings the optimal amplitude is considerably smaller than the amplitude $A \approx 0.73$ required for excitation of light bullets in unmodulated waveguide array. This confirms that in longitudinally modulated structures light bullets can be excited at reduced energy levels. The optimal amplitude grows with increase of frequency detuning [Fig. 2(c)]. Indeed, increasing $|\delta\Omega/\Omega_r|$ results in a stronger diffraction, so that higher peak amplitudes are necessary to counterbalance it and achieve stationary propagation.

An accurate estimate for the optimal amplitude can be obtained for $\Omega = \Omega_r$. Since in this resonant case the diffraction is almost completely suppressed, the light field can be written as $q = w(\eta,\zeta)T(\tau)$, where temporal shape is governed by $iT_\xi = -(\beta/2)T_{\tau\tau} - K|T|^2 T$, where $K = \iint w^4(\eta,\zeta)d\eta d\zeta / \iint w^2(\eta,\zeta)d\eta d\zeta$. The amplitude of light bullet with shape sech($\nu\tau$) is then given by $A = \nu(\beta/K)^{1/2}$. For $\nu = 0.8$ this estimate gives $A = 0.38$ that is only slightly smaller than the optimal amplitude $A_\text{opt} = 0.43$ obtained from direct simulations. Using in Eq. (1) the input wave packets with amplitude $A$ and duration $\nu = A(K/\beta)^{1/2}$ adjusted in order to match this amplitude value, one can show that the range of detunings where effective spatio-temporal localization takes place increases with increase of $A$ [Fig. 2(d)].

The output spatial field distributions (at $\tau = 0$) and temporal dynamics in the central waveguide (at $\eta,\zeta = 0$) are presented in Fig. 3 for the resonant case $\delta\Omega/\Omega_r = 0$ and in Fig. 4 for detuned system with $\delta\Omega/\Omega_r = -0.2$. Figure 3(a) shows that if the input amplitude at $\Omega = \Omega_r$ is too small for the selected input pulse duration one observes pulse spreading and considerable decrease of peak amplitude in the central waveguide. However, it should be stressed that due to the fact that diffraction is almost completely inhibited the most part of the light remains in the central waveguide. Figure 3(b) corresponds to the amplitude providing the maximal value of $U_m$. The formation of light bullet is evident, although one should take into account that the resulting wave packet experiences shape oscillations that are unavoidable just because the refractive index oscillates periodically in neighboring waveguides. In fact temporal dynamics in this case can be improved if one takes somewhat smaller input amplitude (exactly corresponding to the prediction $A = \nu(\beta/K)^{1/2}$ of one-dimensional equation for temporal distribution derived above). Finally, if input amplitude is too high this results in splitting of pulse as shown in Fig. 3(c), although spatial localization remains good. Figure 4 demonstrates the dynamics of bullet excitation in a nonresonant case. The picture in this case is much more complicated. If the input amplitude is not high enough then one observes simultaneous spreading in time and in space [Figs. 4(a) and 4(b)].

If the amplitude is close to the optimal one so that the value of distance-averaged energy is sufficiently high the light bullet forms [Fig. 4(c)]. The corresponding amplitude is notably larger than the amplitude required for bullet formation in a resonant system - a clear evidence of strong effect of diffraction engineering on bullet formation in such structures. It should be noted that similar results were also obtained for different pulse durations and refractive index modulation depths $\mu$.

To illustrate the feasibility of the concept put forward here in current materials with focusing nonlinearity and anomalous GVD, note that similar modulated waveguide arrays made in fused silica have been demonstrated to exhibit the type diffraction control suggested here [19,20,23]. Typical material parameters for such structures are $n_0 = 1.45$, $\partial^2 k / \partial \omega^2 = -280 \text{ fs}^2/\text{cm}$, and $n_2 = 2.2 \times 10^{-16} \text{ cm}^2/\text{W}$ at $\lambda = 1.55 \ \mu\text{m}$. Thus, at $r_0 \sim 30 \ \mu\text{m}$ the width and the separation between waveguides in the array amount to 15 and 60 $\mu\text{m}$, respectively, while $p = 12$ corresponds to a real refractive index modulation depth of $\sim 6 \times 10^{-4}$. Light bullets in such an array should be visible at propagation lengths of the order of 20 cm, when excited with input pulses with a duration $\sim 90$ fs, which corresponds to $L_{\text{disp}} \sim 5.3$ cm, and peak intensities $\sim 40 \text{ GW/cm}^2$, when matching of the effective diffraction, dispersion and self-phase modulation lengths is achieved.

Summarizing, the fundamentally novel result of this Letter is that proper tuning of the longitudinal geometric properties of the material (such as a longitudinal refractive index modulation in a shallow waveguide array imprinted in the material) allows matching the available intrinsic temporal group velocity dispersion and the effective diffraction, thus making possible the generation of stable fully three-dimensional light bullets. Importantly, such bullets are stable in contrast to their counterparts in uniform geometries. They are also excitation-robust. The particular value of the residual or effective diffraction substantially affects the dynamics of light bullet excitation. Consistent with physical expectations, for the resonant modulation frequencies and for the moderate frequency detunings, the optimal input amplitude required for light bullet generation is considerably smaller than in unmodulated arrays. Finally, we stress that the concept put forward here is expected to hold for materials with various nonlinearities and in different refractive index landscapes.

*On leave from Physics Department of M. V. Lomonosov Moscow State University.

# References with titles

**Figure captions**

Figure 1. (a) Refractive index distribution in longitudinally modulated honeycomb array at $\xi = \pi/2\Omega$. Field modulus distribution for a single-site input excitation at $\xi = T_{\mathrm{b}}/2$ in (b) unmodulated array and (c) modulated array with $\Omega = \Omega_{\mathrm{r}}$.

Figure 2. (a) Relative beam spreading at $\xi = 2T_{\mathrm{b}}$ versus frequency detuning. (b) $U_{\mathrm{m}}$ versus $A$ for $\delta\Omega/\Omega_{\mathrm{r}} = 0$ (curve 1) and $-0.2$ (curve 2) at $\nu = 0.8$ and $\xi = 4T_{\mathrm{b}}$. Points in curve 1 correspond to panels (a),(b) in Fig. 3, while points in curve 2 correspond to panels (a)-(c) in Fig. 4. (c) $A_{\mathrm{opt}}$ versus frequency detuning at $\nu = 0.8$. (d) $U_{\mathrm{m}}$ versus frequency detuning for $A = 0.2$ (curve 1) and 0.5 (curve 2) at $\xi = 4T_{\mathrm{b}}$. For each curve the parameter $\nu = A(K/\beta)^{1/2}$.

Figure 3. Field modulus distributions at $\tau = 0$, $\xi = 4T_{\mathrm{b}}$ (top) and temporal dynamics at $\eta, \zeta = 0$ (bottom) for single-site excitations in modulated honeycomb array at $\delta\Omega/\Omega_{\mathrm{r}} = 0$, $\nu = 0.8$ and (a) $A = 0.2$, (b) 0.43, and (c) 0.8.

Figure 4. Field modulus distributions at $\tau = 0$, $\xi = 4T_{\mathrm{b}}$ (top) and temporal dynamics at $\eta, \zeta = 0$ (bottom) for single-site excitations in modulated honeycomb array at $\delta\Omega/\Omega_{\mathrm{r}} = -0.2$, $\nu = 0.8$ and (a) $A = 0.45$, (b) 0.47, and (c) 0.53.

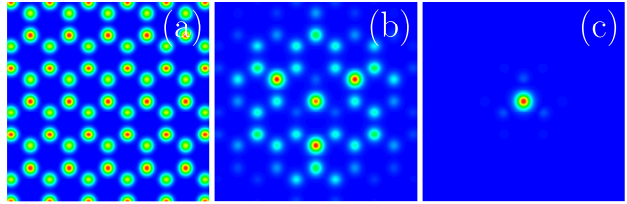

Figure 1.  (a) Refractive index distribution in longitudinally modulated honeycomb array at $\xi = \pi/2\Omega$. Field modulus distribution for a single-site input excitation at $\xi = T_{\rm b}/2$ in (b) unmodulated array and (c) modulated array with $\Omega = \Omega_{\rm r}$.

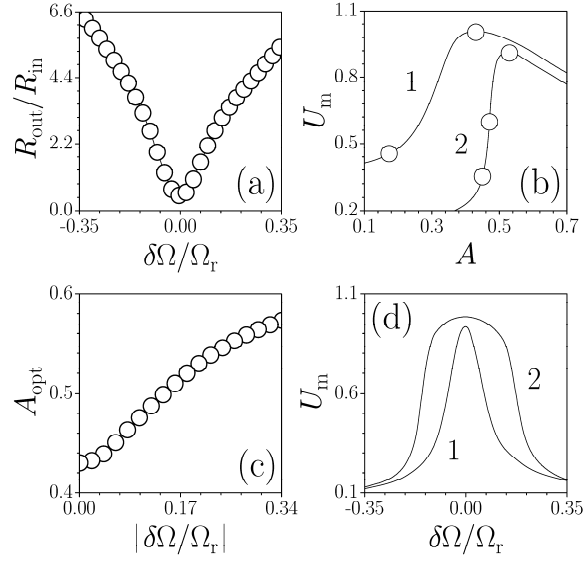

Figure 2. (a) Relative beam spreading at $\xi = 2T_\mathrm{b}$ versus frequency detuning. (b) $U_\mathrm{m}$ versus $A$ for $\delta\Omega/\Omega_\mathrm{r} = 0$ (curve 1) and $-0.2$ (curve 2) at $\nu = 0.8$ and $\xi = 4T_\mathrm{b}$. Points in curve 1 correspond to panels (a),(b) in Fig. 3, while points in curve 2 correspond to panels (a)-(c) in Fig. 4. (c) $A_\mathrm{opt}$ versus frequency detuning at $\nu = 0.8$. (d) $U_\mathrm{m}$ versus frequency detuning for $A = 0.2$ (curve 1) and $0.5$ (curve 2) at $\xi = 4T_\mathrm{b}$. For each curve the parameter $\nu = A(K/\beta)^{1/2}$.

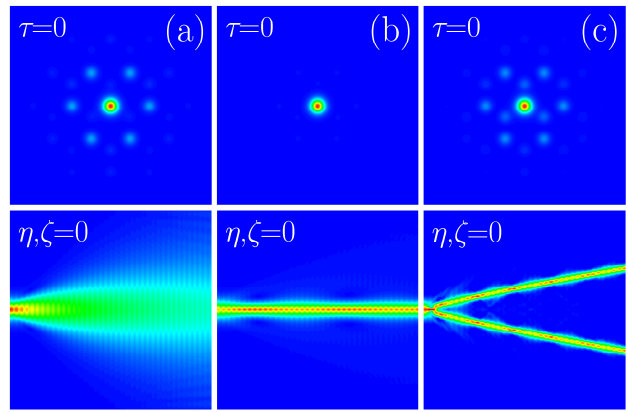

Figure 3. Field modulus distributions at $\tau = 0$, $\xi = 4T_{\rm b}$ (top) and temporal dynamics at $\eta, \zeta = 0$ (bottom) for single-site excitations in modulated honeycomb array at $\delta\Omega/\Omega_{\rm r} = 0$, $\nu = 0.8$ and (a) $A = 0.2$, (b) 0.43, and (c) 0.8.

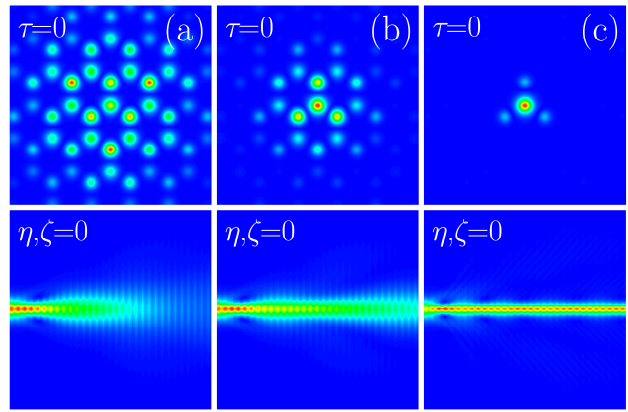

Figure 4. Field modulus distributions at $\tau = 0$, $\xi = 4T_{\rm b}$ (top) and temporal dynamics at $\eta, \zeta = 0$ (bottom) for single-site excitations in modulated honeycomb array at $\delta\Omega/\Omega_{\rm r} = -0.2$, $\nu = 0.8$ and (a) $A = 0.45$, (b) 0.47, and (c) 0.53.